# Comment on the Paper "Stern-Gerlach Force on a Precessing Magnetic Moment" by M. Conte et al[1]

**(Bates 2008 report B/IR#08-01)**

C. Tschalaer

## 1. <u>Introduction</u>

The paper under discussion (ref. 1) is based on an earlier paper (ref. 2). The formula derived for the longitudinal Stern-Gerlach (SG) force in the laboratory frame in ref. 2 (eq. (22)) is not consistent with the total time derivative of the canonical momentum derived from the accepted Langrangian [3] which is (see ref. 4)

$$L = -mc^2/\gamma + e\vec{\beta}\cdot\vec{A} - e\phi + G'/\gamma$$

where

$$G' \equiv \vec{\mu}\cdot\vec{B}'/\gamma + (e/m)(1/(\gamma+1))\vec{s}\cdot(\vec{\beta}\times\vec{E}'/c)$$

is the total SG interaction and Thomas precession potential of a particle with spin $\vec{s}$, magnetic moment $\vec{\mu}$, mass m, and charge e; primed quantities refer to the rest frame of the particle.

The total time derivative of the canonical momentum $\vec{P}$ is thus

$$cd\vec{P}/dt = d[mc^2\gamma\vec{\beta} + e\vec{A} + \delta(G'/\gamma)/\delta\vec{\beta}]/dt$$
$$= c\vec{\nabla}L = ec\vec{\nabla}(\vec{\beta}\cdot\vec{A}) - ec\vec{\nabla}\phi + (c/\gamma)\vec{\nabla}(G').$$

For a particle traversing a localized field region F (all fields and their derivatives are zero outside F), the resulting change $\Delta\vec{p}$ in the momentum $\vec{p}$ is

$$\Delta\vec{p} = \int_F dt \frac{d\vec{p}}{dt} = \int_F dt \frac{d\vec{P}}{dt} = mc\Delta(\gamma\vec{\beta})$$
$$= e\int_F dt\vec{\nabla}[(\vec{\beta}\cdot\vec{A}) - \phi] + (1/\gamma)\int_F (dt/\gamma)\vec{\nabla}(G') \ .$$

The first integral is the electromagnetic interaction and the second the SG-Thomas interaction which contains no terms proportional to $\gamma^2$ and $\gamma$.

## 2. <u>Paper-specific Comment</u>

If, for the sake of discussion, we consider eq. (22) in ref. 2

$$\vec{f}_{SG} = (1/\gamma)\hat{x}\delta(\vec{\mu}\cdot\vec{B}')/\delta x + (1/\gamma)\hat{y}\delta(\vec{\mu}\cdot\vec{B}')/\delta y + \hat{z}\delta(\vec{\mu}\cdot\vec{B}')/\delta z \qquad (22)$$

as the correct expression for the SG force and ignoring the Thomas precession, we can evaluate it using eq. (12) in ref. 2

$$\delta/\delta z' = \gamma\delta/\delta z + (\gamma\beta/c)\delta/\delta t \qquad (12)$$

and by introducing the total time differential

$$d/dt = \delta/\delta t + \beta c \delta/\delta z$$

so that

$$\delta/\delta z' = (\gamma\beta/c)d/dt - \gamma\beta^2\delta/\delta z + \gamma\delta/\delta_z$$
$$= (\gamma\beta/c)d/dt + (1/\gamma)\delta/\delta z \ .$$

Thus eq. (22) yields

$$f_z = (\gamma\beta/c)d(\vec{\mu}\cdot\vec{B}')/dt + (1/\gamma)\vec{\mu}\cdot\delta\vec{B}'/\delta z$$

or

$$dp_z/dt = f_z = mc \ d(\gamma\beta)/dt \ .$$

The total longitudinal momentum change induced by SG forces from a localized field region F is

$$\Delta p_z = \int_F dt \cdot f_z = \int_F dt(\gamma\beta/c) \ d(\vec{\mu}\cdot\vec{B}')/dt + \int_F (dt/\gamma)\vec{\mu}\cdot\delta\vec{B}'/\delta z \ .$$

Using partial integration and remembering that $\vec{B}'$ is zero outside F we find

$$\Delta p_z = -\int_F dt[(\vec{\mu}\cdot\vec{B}'/(mc^2)]d(\gamma\beta mc)/dt \ + \int_F (dt/\gamma)\vec{\mu}\cdot\delta\vec{B}'/\delta z \ .$$

Since

$$\mu mc^2 = (g/2) \ 2.96 \bullet 10^{-11} (MeV)^2/Tesla$$

the SG interaction potential $\vec{\mu}\cdot\vec{B}'$ is much smaller than even the electron rest mass energy i.e.,

$$(\vec{\mu}\cdot\vec{B}')/(mc^2) \cong 1.2 \bullet 10^{-10} B'/Tesla$$

and much smaller for nucleons. Therefore we may approximate

$$\Delta p_z + \int_F dt[\vec{\mu}\cdot\vec{B}'/(mc^2)]dp_z/dt \cong \Delta p_z[1 + (\vec{\mu}\cdot\vec{B}'/(mc^2))_{av}] \cong \Delta p_z$$

or
$$\Delta p_z = \int_F (dt/\gamma) \vec{\mu} \cdot \delta \vec{B}' / \delta_z \times [1 + order(\vec{\mu} \cdot \vec{B}'/(mc^2))]^{-1} ,$$

i.e. there are no first-order $\gamma^2$ – terms in $\Delta p_z$. In ref. 1, this fact is acknowledged for the case of non-precessing spin. However, it is then claimed that, for a particular model with precessing spin, the $\gamma^2$ term survives. We now show that if the spin motion is accounted for properly, the $\gamma^2$ terms cancel again as in the case of fixed spin.

## 3. **Precessing spin in a $T_{011}$ field**

A particle traveling on axis through a rectangular RF cavity ($x = a/2$ ; $y = b/2$) containing a $T_{011}$ field as described in ref. 1 "sees" the following RF field $\vec{B}'$:

$$B'_x = \gamma B_x + \gamma \beta E_y / c = 0$$
$$B'_y = \gamma B_y - \gamma \beta E_x / c = -\gamma B_0 (b/d)[\cos(\pi \zeta) \cos \omega t - w\beta \sin(\pi \zeta) \sin \omega t]$$
$$B'_z = B_z = 0$$

where
$$\zeta \equiv z/d$$
$$\omega t = \pi w \zeta / \beta + \varphi$$
$$w \equiv \sqrt{1 + d^2/b^2} = \omega d/(\pi c) .$$

Eqs. (12) and (22) then yield
$$f_z = (\gamma \beta / c) \delta(\vec{\mu} \cdot \vec{B}') / \delta t + \gamma \vec{\mu} \cdot \delta \vec{B}' / \delta z .$$

Spin rotation in a magnetic field is described by [5]
$$\delta \vec{\mu} / \delta t = \vec{\mu} \times \vec{B}' \cdot eg/(2m) .$$

Therefore, $\vec{B}' \cdot \delta \vec{\mu} / \delta t$ is zero and
$$f_z = (\gamma \beta / c) \vec{\mu} \cdot \delta \vec{B}' / \delta t + \gamma \vec{\mu} \cdot \delta \vec{B}' / \delta z$$
$$= \gamma \mu_y [\delta B'_y / \delta z + (\beta/c) \delta B'_y / \delta t] .$$

If we now apply a strong constant field $B'_x \gg B'_y$ as suggested in ref. 1 to precess the moment $\vec{\mu}$ which points in the z-direction on entering the cavity, the spin rotation equations yields

$$d\mu_z / dt = (\mu_x B'_y - \mu_y B'_x) eg/(2m\gamma) \cong -\mu_y B'_x \, eg/(2m\gamma) = -\Omega \mu_y$$
$$d\mu_y / dt = \mu_z B'_x \, eg/(2m\gamma) = \Omega \mu_z$$

where
$$\Omega \equiv B'_x \, eg/(2m\gamma)$$

which has the solutions

$$\mu_y = \mu \sin[\Omega(t+t_0)]; \quad \mu_z = \mu \cos[\Omega(t-t_0)] .$$

In ref. 1., $B'_x$ was chosen to produce a 180° rotation of $\vec{\mu}$ in a cavity traversal i.e.

$$\Omega(t-t_0) = \pi_\zeta$$

and

$$B'_x \, eg/(2m\gamma) = \pi\beta c/d$$

However, spin rotation also results in a component $\mu_x$ produced by $B'_y$

$$d\mu_x/dt = -\mu_z B'_y \, eg/(2m\gamma) = -\pi\beta c \mu B'_y/B'_x \cdot \cos(\pi\zeta) .$$

After exiting the cavity, this moment $\mu_x$

$$\mu_x = -\pi\beta c \mu/B'_x \int_F dt \, \cos(\pi\zeta) B'_y$$

will generate a SG force in the fringe field $B'_x$ according to eq. (22):

$$dp_z/dt = \gamma \mu_x \delta B_x' / \delta z$$

which produces a longitudinal momentum change $\Delta p_z$ in traversing the $B_x'$ fringe field

$$\Delta p_2 = \mu_x \gamma/(bc) \cdot \int_{fringe} dz \, \delta B'_x/\delta z = -\mu_x B'_x \gamma/(\beta c)$$
$$= \gamma \mu \pi \int_F dt \, \cos(\pi\zeta) B'_y .$$

The longitudinal momentum change $\Delta p_1$ produced by the SG force $f_z$ in traversing the cavity is

$$\Delta p_1 = \gamma \mu \int_F dt \, \sin(\pi\zeta)[\delta B'_y/\delta z + (\beta/c)\delta B'_y/\delta t] .$$

The integrals for $\Delta p_1$ and $\Delta p_2$ can now be evaluated using

$$dt = d\zeta \cdot d/(\beta c) .$$

We find

$$\delta B'_y/\delta z + (\beta/c)\delta B'_y/\delta t = \gamma \pi B_0 b/d^2 \cdot [(1+w^2\beta^2)\sin(\pi\zeta)\cos\omega t + 2w\beta\cos(\pi\zeta)\sin\omega t]$$

and

$$\Delta p_1 = \gamma^2 \pi \mu B_0 b/(\beta cd) \cdot [(1+\beta^2 w^2)\int_0^1 d\zeta \frac{1-\cos(2\pi\zeta)}{2}\cos(\omega t) + 2w\beta \int_0^1 \frac{\sin(2\pi\zeta)}{2}\sin(\omega t)]$$

$$\Delta p_2 = -\gamma^2 \pi \mu B_0 b/(\beta cd)[\int_0^1 d\zeta \frac{1+\cos(2\pi\zeta)}{2}\cos(\omega t) - w\beta \int_0^1 d\zeta \frac{\sin(2\pi\zeta)}{2}\sin(\omega t)] .$$

The total longitudinal momentum change is therefore

$$\Delta p_1 + \Delta p_2 = \gamma^2 \mu B_0 b/(2\beta cd)[\beta^2 w^2 I_0 - (2+\beta^2 w^2)I_c + 3w\beta I_s]$$

where

$$I_0 \equiv \pi \int_0^1 d\zeta \cos(\pi w\zeta/\beta + \varphi) = (\beta/w)[\sin(\pi w/\beta + \varphi) - \sin\varphi]$$
$$I_c \equiv \pi \int_0^1 d\zeta \cos(2\pi\zeta)\cos(\omega t) = (w/\beta)/(w^2/\beta^2 - 4) \cdot [\sin(\pi w/\beta + \varphi) - \sin\varphi]$$
$$I_s \equiv \pi \int_0^1 d\zeta \sin(2\pi\zeta)\sin(\omega t) = 2/(w^2/\beta^2 - 4) \cdot [\sin(\pi w/\beta + \varphi) - \sin\varphi].$$

Thus

$$\Delta p_1 + \Delta p_2 = [\gamma^2 \mu B_0 b/(2\beta cd)](2w/\beta)[(2\beta^2 - 1)/\gamma^2][\sin(\pi w/\beta + \varphi) - \sin\varphi]$$
$$= [\mu B_0(b/d)w(2\beta^2 - 1)/(\beta^2 c)][\sin(\pi w/\beta + \varphi) - \sin\varphi].$$

Therefore, the $\gamma^2$ terms of the longitudinal momentum change caused by the SG forces from the RF fields and by the fall-off of the precessing field $B_x'$ cancel as predicted by the more general formalism.

The use of two equal cavities placed in series and containing RF fields of the same mode as was proposed in Ref. 1 simply extends the integrals $I_o$, $I_c$, and $I_s$ from 0 to 2 instead of from 0 to 1. Then the term sin ($\pi w/\beta+\varphi$) in the last equation is replaced by sin ($2\pi w/\beta+ \varphi$), but this does not affect the cancellation of the $\gamma^2$ terms.

## **Conclusion**

In ref. 1 it is claimed that the SG interaction results in a change of the longitudinal momentum proportional to $\gamma^2$ for a particle traversing a localized electromagnetic field and that this should be investigated as an effective way to polarize stored antiproton beams. However, this claim is inconsistent with predictions of refs. 3 and 4, from which it can be shown that such $\gamma^2$-terms are absent.

In this paper, we have shown that the terms in $\gamma^2$ in ref. 1 are cancelled once the complete spin motion is included, thereby removing that contradiction with ref. 4.

It is worth noting that the Thomas precession included in ref. 4 does not change this picture but only modifies the strength of the non-cancelling terms.